\documentclass{ptephy_v1}

\preprintnumber{XXXX-XXXX} 

\usepackage{graphicx}
\graphicspath{{./image/}}
\usepackage{subfig}
\usepackage[separate-uncertainty=true,multi-part-units=single]{siunitx}

\begin{document}

\title{The development of super fine-grained nuclear emulsion}


\author{Takashi Asada}
\affil{Graduate School of Science, Nagoya University, Furo-cho, Chikusa-ku, Nagoya, 464-8602, Japan \email{asada@flab.phys.nagoya-u.ac.jp}}

\author{Tatsuhiro Naka}
\affil{Kobayashi-Maskawa Institute for the Origin of Particles and the Universe, Center for Experimental Studies, Nagoya University, Furou-cho, Chigusa-ku, Nagoya, 464-8602, Japan}

\author[1]{Ken-ichi Kuwabara}
\author[1]{Masahiro Yoshimoto} 


\begin{abstract}%
A nuclear emulsion with micronized crystals is required for the tracking detection of submicron ionizing particles, which are a target of dark matter detection and other methods. We found that a new production method, named as the PVA-Gelatin Mixing Method (PGMM), could effectively control crystal size from 20 nm to 50 nm. We named two types of an emulsion produced with the new method NIT and UNIT. The composition and spatial resolution of them were measured, and the results indicated that these emulsions detect extremely short tracks.
\end{abstract}

\subjectindex{H20, H21}

\maketitle

\section{Introduction}
A nuclear emulsion is a solid-state tracking detector with extremely high spatial resolution used in particle detectors. Micronized emulsions enable the direction of low-velocity ionizing particles to be detected, even at submicron lengths. These emulsions are thus useful in a wide range of applications, such as neutrino coherent scattering~\cite{1}, fast neutron detection~\cite{bibi_MevNeutron}, and directional dark matter searches~\cite{bibi_directional}. The mass scalability and spatial resolution of emulsions are of great benefit to directional dark matter search experiments~\cite{2}. Moreover, micronized emulsions are useful for the detection of high-energy heavy ions because of their wide dynamic range for charge discrimination.

	The spatial resolution of an emulsion is determined by the size of its silver halide crystals. As an ionizing particle passes through an emulsion, the particle ionizes crystals along its path and ionized electrons reduce several silver ions to silver atoms which become the foundation of the signal. This process is confined within each crystal; the position accuracy of signals is thus dependent on crystal size.
	
	The direction of a track is detected when the charged particle passes through at least two crystals. The detectable track range thus depends on the average distance of the crystals, and this distance depends mainly on crystal size~\cite{3}. Therefore, smaller crystals are required for shorter track detection.

	Earlier studies on short-track detection were performed with an emulsion containing crystals of $\SI{40}{nm}$~\cite{3}. However, this emulsion was unstable, and its crystals often agglutinated and grew. We studied methods of creating smaller crystals by constructing an emulsion production machine at Nagoya University~\cite{2} and succeeded in controlling the crystal size.

	In this paper, we describe a new emulsion production method using a mixed solution of gelatin and polyvinyl alcohol (PVA). This approach stably provides micronized crystals. Moreover, we report measurements of crystal size and composition for the emulsion, as well as its theoretical spatial resolution.

\section{Production of emulsions with micronized silver halide crystals}
\subsection{Overview of production method}
 	The primary material of a nuclear emulsion is a silver halide emulsion. In an emulsion, the silver halide crystals are uniformly dispersed in a binder such as gelatin. The ideal crystals for a nuclear emulsion are monodispersed, spherical, and micronized. 

\begin{figure}[htbp]
 \begin{center}
  \includegraphics[width=0.9\columnwidth,clip]{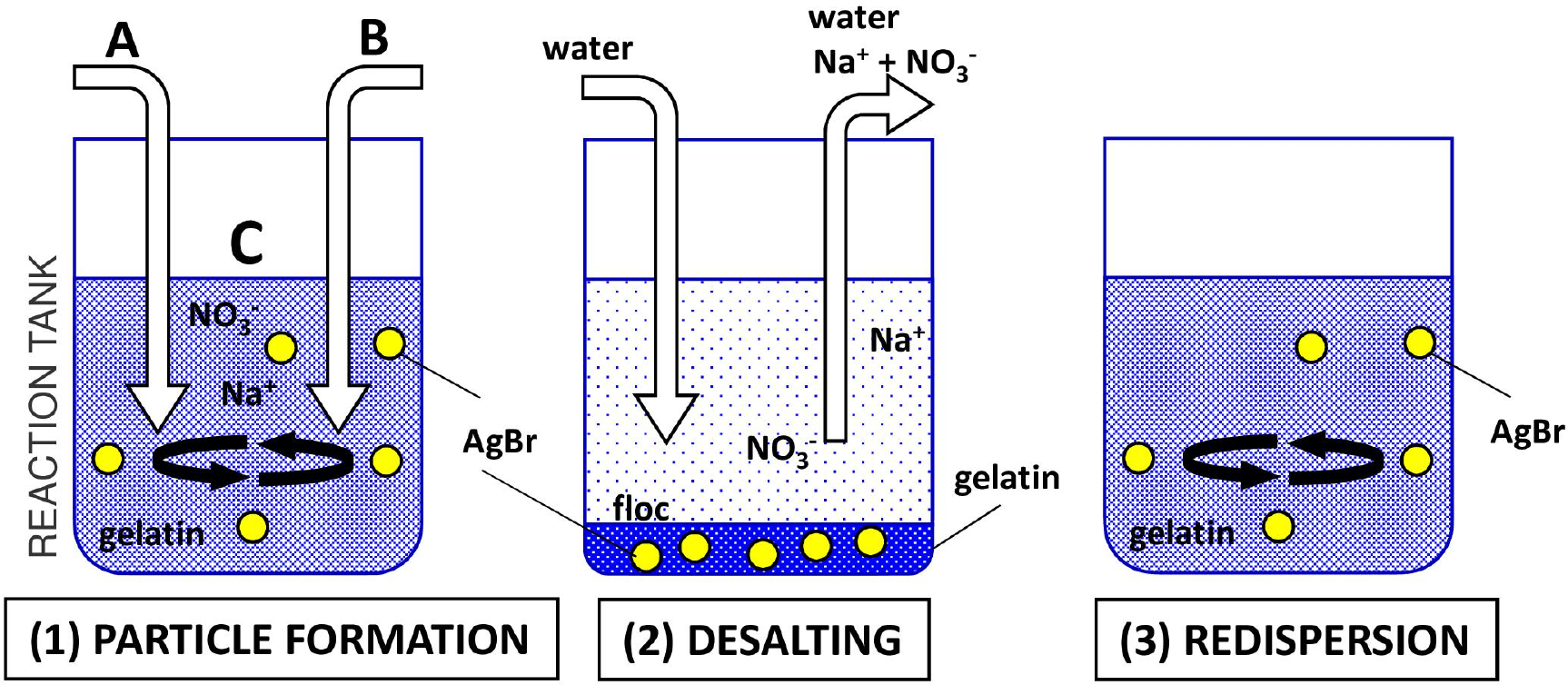}
  \caption{Diagram of emulsion production steps. The steps proceed from left to right: (1) particle formation, (2) desalting, and (3) redispersion.}
  \label{figure1}
 \end{center}
\end{figure}

	We installed an emulsion production facility to promote the research and development of the ideal nuclear emulsion. The production capacity of our installation is on the order of $\SI{100}{g}$ per batch under dry conditions, with two batches per day. Emulsion production requires three steps: particle formation, desalting, and redispersion (Fig.~\ref{figure1}). Our new method proceeds through these steps in the presence of a PVA solution. 

	Some polymers have been reported to strongly suppress the growth of silver halide crystals~\cite{4}~\cite{5}. This effect is also maintained in mixed solutions with another polymer. However, these polymers have handling problems when serving as binders. Therefore, gelatin is the current best candidate for a binder.

	In the new method, we produced an emulsion by a typical method using a gelatin solution, but the solution contained PVA exceeding $\SI{0.25}{wt\%}$ to avoid crystal agglutination and growth. We used an alkali-treated gelatin derived from cattle bone, P6405 produced by Nitta Gelatin Inc., in the present study. The isoelectric point of the gelatin was 4.8--5.2. We chose PVA with a molecular weight of 89000--98000 to avoid its unexpected dissolving.

\subsection{Particle formation}\label{sec_partice_formation}
Particle formation is the most important step for producing silver halide crystals by chemical reaction and controlling their size and form. We prepared a reaction tank with a constant temperature and rotational speed. The tank held a solution of gelatin and PVA (C), then solutions of silver nitrate (A) and halogen (B) were added into the tank at the same time at constant speed. The silver and halogen generated crystals as sediment through the reaction given in formula (\ref{reaction}), and the crystals were kept dispersed by a protective colloid of gelatin and PVA.
  
\begin{equation}\label{reaction}
\mathrm{Ag^+} + \mathrm{X^-} \rightarrow \mathrm{AgX}
\end{equation}

 We used $\SI{340}{mL}$ of solution A, consisting of $\SI{0.427}{mol.L^{-1}}$ of AgNO$_3$ and $\SI{0.119}{mol.L^{-1}}$ NH$_4$NO$_3$. We also used $\SI{340}{mL}$ of solution B, consisting of $\SI{0.420}{mol.L^{-1}}$ of NaBr (or KBr) and $\SI{0.015}{mol.L^{-1}}$ NaI (or KI). Solution C is consisted of $\SI{29}{g}$ gelatin dissolved in $\SI{1000}{mL}$ water at $\SI{40}{\degreeCelsius}$ and $\SI{5}{g}$ PVA dissolved in $\SI{300}{mL}$ water at $\SI{80}{\degreeCelsius}$.
 
 We kept solution C at $\SI{35.0 \pm 0.1}{\degreeCelsius}$ and a rotation rate of $\SI{1000}{rpm}$ and added $\SI{2e-5}{mol}$ of acid sodium salt. The addition of solutions A and B to solution C produced silver iodobromide crystals containing $\SI{3.5}{mol\%}$ under the above conditions at addition speeds of $\SI{57 \pm 1}{mL.min^{-1}}$ and $\SI{10 \pm 1}{mL.min^{-1}}$, correspond to addition times of 6 min and 35 min, respectively. The concentration of iodine was not critical for crystal size control, but it affects sensitivity, and nuclear emulsions typically use iodobromide.

\subsection{Desalting}
	Sodium and nitrate ions remained as salts in the crystal formation step. The desalting step, also called the washing step, removes these salts. We used the flocculation method, which is habitually used in photographic emulsion production. First, we added a polyvalent anionic compound of carboxylic acid called the flocculation agent to the emulsion made by the particle formation step. The gelatin including the silver halide crystals coagulated and formed a sediment called floc when the pH was adjusted to $3.6$ with sulfuric acid. The remaining salt was then removed as the supernatant liquid. This process is called coagulating sedimentation or flocculation. We repeated the process by adding new water until the electric conductivity of supernatant liquid decreased below $\SI{1}{mS.cm^{-1}}$, indicating that the remaining salt was $1/100$ its beginning amount.

	In the desalting step, gelatin loses its protective colloid properties, which often leads to the agglutination and growth of crystals. Additionally, the flocculation agent could not precipitate PVA in the coagulating sedimentation process, which reduced the PVA concentration. Silver halide crystals occasionally agglutinated and grew in the low-concentration PVA solution in this step. We used $\SI{0.25}{wt\%}$ PVA solution instead of water alone to maintain the PVA concentration and avoid this process. We succeeded at suppressing unexpected crystal growth in this step with this method.

\subsection{Redispersion}
	We obtained a uniformly dispersed emulsion through the redispersion step. The floc from the desalting step had remained as precipitate. We controlled the pH of the solution at pH $5.5$--$6.0$ using sodium hydroxide, which solved the precipitation state. We then kept the solution at $\SI{50 \pm 1}{\degreeCelsius}$ and a rotation rate of $\SI{1000}{rpm}$ for at least 1 hour to uniformly disperse the gelatin and crystals. The final product became the material of the nuclear emulsion.

\subsection{Summary of production method}
 We developed a new production method for micronized nuclear emulsions. We used a gelatin-PVA mixed solution in the particle formation step, and the diluent in the desalting step was a PVA solution instead of water. We named this production approach the PVA-Gelatin Mixing Method (PGMM). This process yielded stable micronized crystals, as reported by the following section.

\section{Evaluation}
\begin{figure}[htbp]
 \begin{center}
  \subfloat[]{
   \includegraphics[width=0.49\columnwidth,clip]{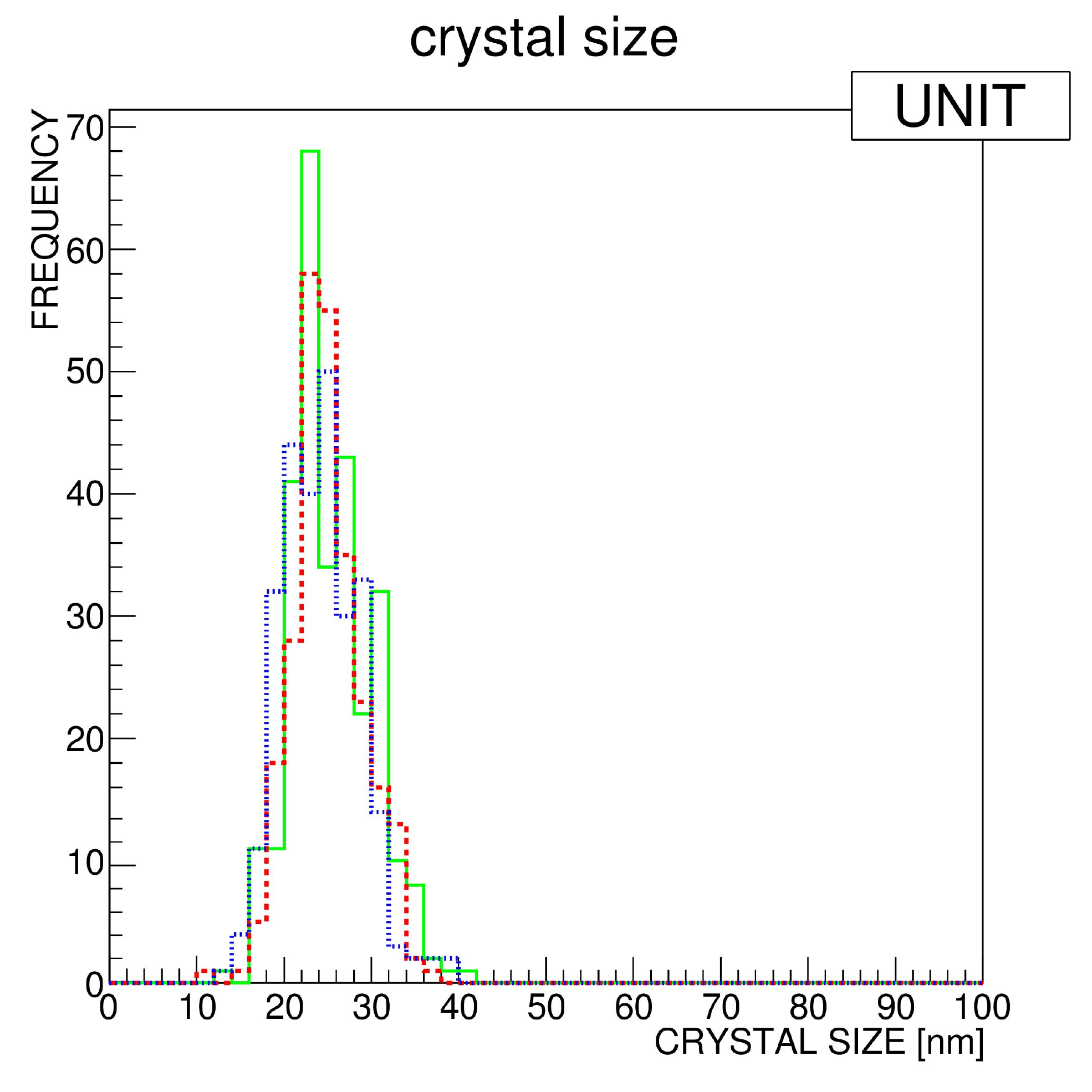}
  }
  \subfloat[]{
   \includegraphics[width=0.49\columnwidth,clip]{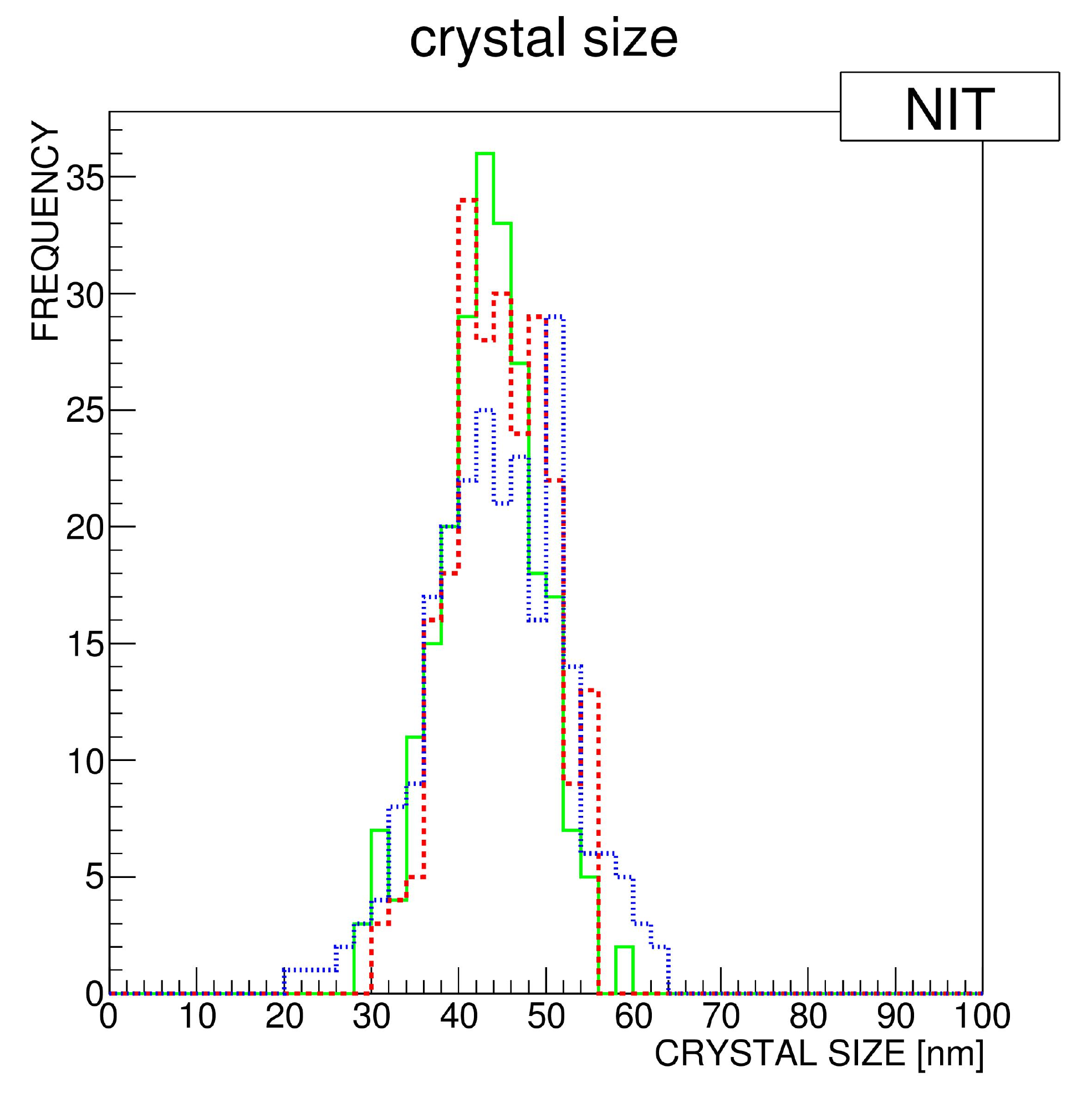}
  }
  \caption{Dispersion of crystal diameter. (a), UNIT; (b), NIT. Each plot represents three batches with each different color.}
  \label{figure2}
 \end{center} 
\end{figure}

 The diameter of silver halide crystals in the binder was measured by transmission electron microscopy (TEM). We prepared spread crystals diluted with aqueous gelatin solution. We took TEM images on a JEOL JEM-1400EX and measured the diameter of the crystals. Fig.~\ref{figure2} plots the diameter distribution of crystals made with addition speeds of $\SI{57}{mL.min^{-1}}$~(a) and $\SI{10}{mL.min^{-1}}$~(b) explained in \ref{sec_partice_formation}. The average length of crystals in (a) is $\SI{24.8 \pm 0.1}{nm}$, with a standard deviation of $\SI{4.3}{nm}$, and that of (b) is $\SI{44.2 \pm 0.2}{nm}$, with a standard deviation of $\SI{6.8}{nm}$. These plots are derived from the mixing data for three batches. We designated (a) as Ultra Nano Imaging Tracker (UNIT) and (b) as Nano Imaging Tracker (NIT).

\begin{figure}[htbp]
 \begin{center}
  \includegraphics[width=0.7\columnwidth,clip]{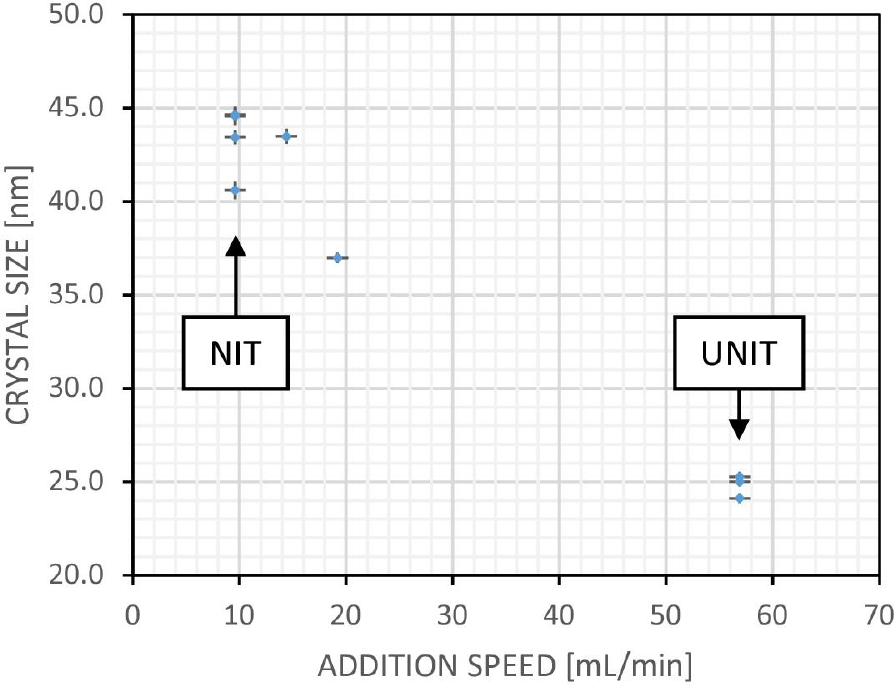}
   \caption{Relationship of addition speed crystal size.}
   \label{figure3}
 \end{center}
\end{figure}

 Fig.~\ref{figure3} shows the relationship between addition speed and measured crystal diameter. Addition speed and crystal size are inversely correlated, and the size varied by less than $\SI{5}{nm}$ at the same addition speed. This plot indicates that crystal size was controlled by addition speed.
 
 It is empirically known that using a same addition time reproduces the crystal size in other scales of production volume because it gives same condition of concentration in the chemical reaction. The times are $\SI{6}{min}$ for UNIT and $\SI{35}{min}$ for NIT calculated from addition speed and addition volume described in \ref{sec_partice_formation}.

\subsection{Elemental composition and density}
The characteristics as a target, energy deposit of a passing particle and cross section for detection of a neutral particle, depend on the elemental composition and density.  First, we determined the elemental composition of a dry emulsion used under experimental conditions. We assumed that the ratio of silver to halogen was one to one because they are from a slightly soluble salt and the desalting process removed the excess ions. Only the bromide ion was removed in this assumption because the solubility product of AgBr ($\SI{8.5e-13}{mol^2.L^{-2}}$) was larger than that of AgI ($\SI{1.74e-16}{mol^2.L^{-2}}$) by nearly four order of magnitude in the desalting process. We confirmed that the percentages of silver and bromine calculated from the raw materials were consistent within $2\%$ with the results of Scanning Electron Microscopy with Energy Dispersive X-Ray Spectroscopy (SEM-EDX). The amount of iodine was less than the detection limit. Therefore, we adopted the estimated ratio for silver, bromine, and iodine (Table~\ref{tb_RawMterials}\subref{tb_AgBrI}).

 Next, we considered light atoms except silver and halogen. The percentages of gelatin and PVA usually change during the desalting process. Therefore, we calculate the ratio from the absolute mass fractions of carbon, hydrogen, and nitrogen directly measured by a YANACO MT-6 elemental analyzer. The absolute accuracy was $0.3\%$, and the relative humidity in the measurement room was $30\%$--$40\%$.

 The ratios of elements were calculated as follows. We ignored additional ingredients present at less than $0.3\%$ of the mass fraction of the raw materials and considered the binder to be a pure mixture of gelatin and PVA. We measured the elemental fractions of gelatin (Table~\ref{tb_RawMterials}\subref{tb_gelatin}), PVA (Table~\ref{tb_RawMterials}\subref{tb_PVA}), and produced emulsion by the elemental analyzer. We determined the ratio of gelatin in emulsion from the mass fraction of nitrogen, because PVA do not have nitrogen and the nitrogen purely came from gelatin. Hydrogen and carbon except these of gelatin were from PVA. The elements of remained parts of emulsion were silver, bromine, and iodine.
 
 Table~\ref{tb_emulsion} shows the estimated mass fraction and atomic fraction of emulsion. The emulsion density is $\SI{3.44}{g.cm^{-3}}$ in the dry condition, summing up densities of $\SI{6.473}{g.cm^{-3}}$, $\SI{1.32}{g.cm^{-3}}$ and $\SI{1.19}{g.cm^{-3}}$ and mass fractions of $78.1\%$, $16.8\%$, and $5.1\%$ for silver halide, gelatin, and PVA, respectively.

Furthermore, we cross-checked the mass ratio of silver halide to binder by another method. Photographic fixing processing removes only silver halide from an emulsion. Therefore, we measured the mass ratio by the mass difference before and after this processing. We obtained the mass ratio of $\SI{76.3 \pm 0.6}{\%}$ from fixing processing under relative humidity of $60\%$--$70\%$. The value is almost consistent to $\SI{78.1\pm 0.8}{\%}$, calculated from Table~\ref{tb_emulsion}.

The water-absorbing properties of gelatin and PVA changes the water content in the emulsion depending on relative humidity. The difference is usually about $1\%$ in maximum. The water content should be corrected considering the relative humidity of the local environment of the experiment. In particular, the humidity modifies hydrogen ratio which directly affects to neutron recoil events. When the water decreases by $1\%$ of emulsion weight, the hydrogen decreases by $8\%$ of its atomic fraction.

\begin{table}[htbp]
\begin{center}
\caption{The mass fractions and atomic fractions of constituent of the emulsion.}
\label{tb_RawMterials}

\subfloat[The mass fraction and atomic fraction of silver halide estimated from the rate of the raw materials]{
\label{tb_AgBrI}
\begin{tabular}{ccc} 
\hline
Element & Mass$\%$ & Atom$\%$ \\
\hline
Ag & 56.9 & 50.0 \\
Br & 40.6 & 48.2 \\
I & 2.4 & 1.8 \\
\hline
\multicolumn{3}{c}{} \\
\end{tabular}
}
\\
\begin{tabular}{cc}
\subfloat[The mass fraction and atomic fraction of gelatin measured by the elemental analyzer]{
\label{tb_gelatin}
\begin{tabular}{ccc} 

\hline
Element & Mass$\%$ & Atom$\%$ \\
\hline
C & 43.7 & 26.1 \\
H & 7.1 & 50.8 \\
N & 16.0 & 8.2 \\
O & 33.1 & 14.8 \\
\hline
\end{tabular}
}
\hspace{2cm}
\subfloat[The mass fraction and atomic fraction of PVA estimated from the elemental composition]{
\label{tb_PVA}
\begin{tabular}{ccc} 

\hline
Element & Mass$\%$ & Atom$\%$ \\
\hline
C & 54.5 & 28.6 \\
H & 9.1 & 57.1 \\
O & 36.4 & 14.3 \\
\hline
\multicolumn{3}{c}{} \\
\end{tabular}
}
\end{tabular}

\end{center}

\end{table}

\begin{table}[htbp]
  \centering
    \caption{Estimated mass fraction and atomic fraction of produced emulsion.}
    \label{tb_emulsion}
    \begin{tabular}{ccc}
    \hline
    Element & Mass$\%$ & Atom$\%$ \\
    \hline
    Ag    & 44.5  & 10.5 \\
    Br    & 31.8  & 10.1 \\
    I     & 1.9  & 0.4 \\
    C     & 10.1  & 21.4 \\
    N     & 2.7   & 4.9 \\
    O     & 7.4   & 11.7 \\
    H     & 1.6   & 41.1 \\
    \hline
    \end{tabular}%
\end{table}%

\subsection{Spatial resolution}
    The volume-filling factor of silver halide crystals is $41.5\%$ due to the mass fraction of silver halide. The average number of crystals per length was 25 per $\si{\micro m}$ for UNIT and 14 per $\si{\micro m}$ for NIT assuming spherical crystals and each crystal size. The average distance between crystals is reciprocal to this value and was thus $\SI{40}{nm}$ for UNIT and $\SI{71}{nm}$ for UNIT. These values correspond to the principal spatial resolution when the crystals are used in a tracking detector. The average distance is equal to the detection limit of track length if the ``crystal sensitivity'', the reaction rate of each crystal passed by ionizing particle, is $100\%$. 

 In this case, the average distance is converted to detectable particle energy by SRIM~\cite{6} which is software calculating transport of ions in matter using the measured components and density as follows: $\SI{13}{keV}$ on UNIT and $\SI{25}{keV}$ on NIT for carbon, $\SI{55}{keV}$ on UNIT and $\SI{160}{keV}$ on NIT for silver. The effective detection threshold depends on the crystal sensitivity for each particle.

\section{Performance control}

\begin{figure}[httb]
 \begin{center}
  \subfloat[NIT with HA]{
   \includegraphics[width=0.45\columnwidth,clip]{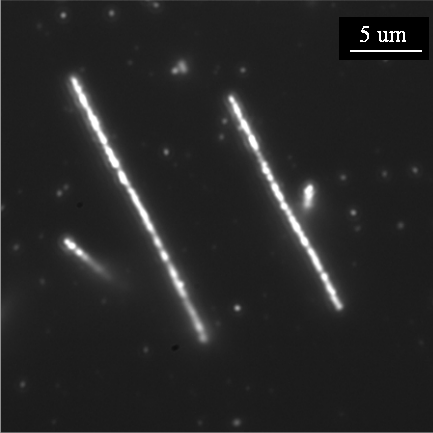}
   \label{subfig_NITHA}
  }
  \subfloat[UNIT with HA]{
   \includegraphics[width=0.45\columnwidth,clip]{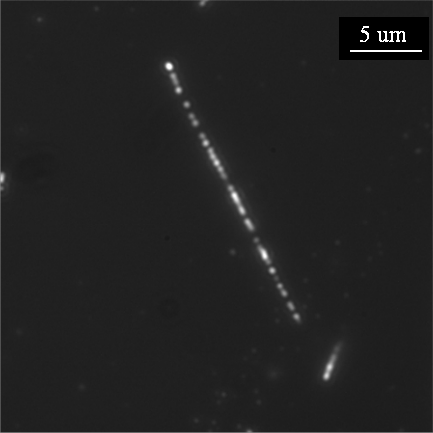}
   \label{subfig_UNITHA}
  }
  \hspace{0mm}
  \subfloat[NIT without HA]{
   \includegraphics[width=0.45\columnwidth,clip]{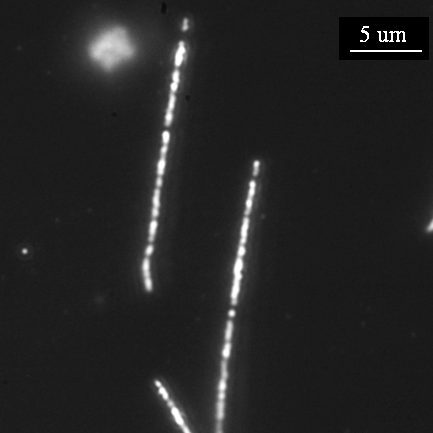}
   \label{subfig_NIT}
  }
  \subfloat[UNIT without HA]{
   \includegraphics[width=0.45\columnwidth,clip]{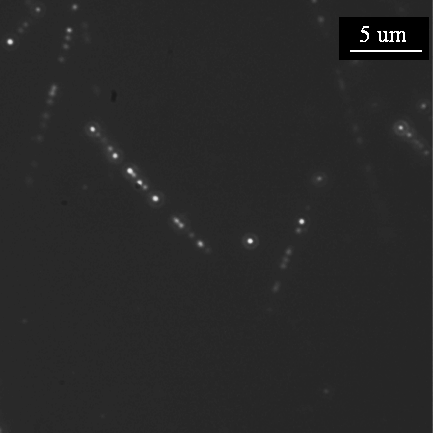}
   \label{subfig_UNIT}
  }
  \caption{Images of alpha-ray tracks in each emulsion under an optical microscope. Alpha rays were induced by $^{241}$Am. The emulsions were developed by a developer for OPERA experiment~\cite{8}.}
  \label{figure4}
 \end{center} 
\end{figure}

 Various factors affect the sensitivity of nuclear emulsions. The nuclear emulsion usually loses signal as time passes after recording, which is called fading. The quantity of ionization depends on the passing length of a particle in a crystal. Chemical sensitization and the developing process also modify the sensitivity of each crystal.

 Sensitivity requirements vary due to the kind of passing particle and its energy deposit. The crystal sensitivity also depends on experimental conditions such as temperature, humidity, and the presence of oxygen. Therefore, users should adjust the sensitivity of the crystal for each experiment.

 Fig.~\ref{figure4} shows examples of sensitivity control by crystal size and chemical sensitization. The examples here use halogen acceptor sensitization (HA)~\cite{7}, a standard chemical sensitization method for micronized crystals. 

 A line of white dots represents one track from one alpha ray, and one dot is called a grain, which is the silver cluster developed from one crystal. If the crystal sensitivity is low, the number of grains per length is sparser comparing to the number of crystal per length. If all of the interacted crystals are developed, the nuclear emulsion reaches its theoretical spatial resolution.

 However, grain size and the possibility of development depend on crystal size, sensitization treatment, developing treatment, and the energy deposition of the target. Against the tracking detection limit decided from the grain density per length, the resolution will be restricted by readout efficiency, which depends on grain size or background events due to the S/N ratio. The best conditions differ for each experiment. In general, smaller crystal size reduces crystal sensitivity and grain size but increases total crystal amount per length. Stronger sensitization and developing treatment give higher sensitivity and noise.

 In Fig.~\ref{figure4}, NIT\subref{subfig_NITHA} with HA and \subref{subfig_UNITHA}~UNIT with HA have almost the same grain density per length against their grain sizes. These pictures indicate that \subref{subfig_UNITHA}~UNIT with HA is roughly half as sensitive as \subref{subfig_NITHA}~NIT with HA for alpha rays. \subref{subfig_NITHA}~NIT with HA is thus best for alpha-ray detection because of its large grain and good grain density. If particle discrimination by energy deposition is necessary, UNIT \subref{subfig_UNITHA}~with or \subref{subfig_UNIT}~without HA is better because of its upper margin of sensitivity, which provides a wide dynamic range. Superior spatial resolution requires higher grain density and small grain size; therefore, \subref{subfig_UNITHA}~UNIT with HA or further sensitization is appropriate for pursuance of spatial resolution.

\begin{figure}[httb]
    \begin{center}
        \includegraphics[width=0.7\columnwidth,clip]{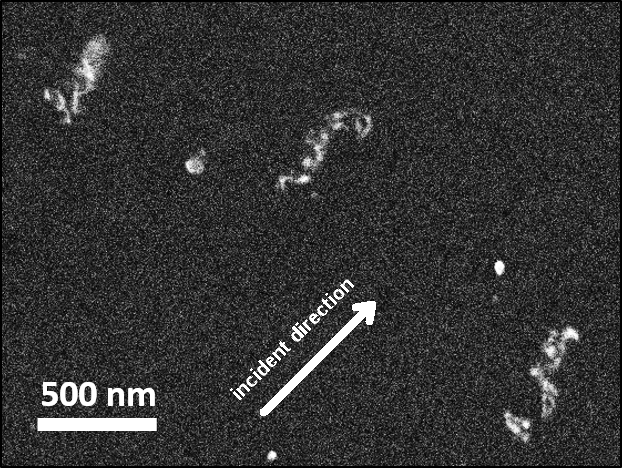}
            \caption{Image of carbon tracks under SEM. Carbon was induced at $\SI{100}{keV}$ by an ion implantation system in the direction shown by an arrow. The emulsion is NIT with HA and was developed by a Metol ascorbic acid (MAA) developer~\cite{9}.}
            \label{figure5}
    \end{center}
\end{figure}

Fig.~\ref{figure5} shows an actual image of extremely short tracks taken by Scanning Electron Microscopy (SEM). A $\SI{100}{keV}$ carbon ion was dosed to NIT with HA by an ion implantation machine. A track has grains of filamented shape with the resolution of SEM. The three tracks in the picture indicate the induced direction by their similar angles. The interval between grains is around 100 nm, which implies that the crystal sensitivity is close to $100\%$. Further fine tuning of tracks is possible, as illustrated in Fig.~\ref{figure4}.

 This nuclear emulsion has many applications in practical usage. The crystal size, sensitization treatment, and developing treatment should be adjusted to the purpose of each experiment, as mentioned above. The phenomenology of the particle detection mechanism for fine silver halide crystals has already been discussed in previous reports~\cite{10}.

\section{Conclusion}
 We succeeded in developing a nuclear emulsion with highly micronized crystals that allows extremely short track detection. Our new production method, known as PGMM, stably produces $20$--$\SI{50}{nm}$ crystals. These sizes depend on the addition speed during the particle formation process and are controllable. We designated $\SI{44.2 \pm 0.2}{nm}$ crystals as NIT and $\SI{24.8 \pm 0.1}{nm}$ crystals as UNIT. 

 The spatial resolution corresponds to the average distance between the crystals. These distances are $\SI{71}{nm}$ for NIT and $\SI{40}{nm}$ for UNIT.  Therefore, tracks of these ranges are detectable, assuming a detection efficiency of $100\%$. These ranges are equivalent to $\SI{13}{keV}$ for NIT and $\SI{25}{keV}$ for UNIT for the carbon track case. The directional sensitivity for low-energy atoms is very important for performing several unique experiments.

\section*{Acknowledgment}
    We thank K. Oyama of the Chemical Instrumentation Facility, Nagoya University, Japan for the measurements of the nuclear emulsion composition. The gelatin used in this study was provided by Nitta Gelatin Inc.
    
 This work was supported by a Grant-in-Aid for JSPS Research Fellows, Ministry of Education, Culture, Sports, Science and Technology of Japan, and a JSPS Grant-in-Aid for Young Scientists (A) (15H05446) and Grant-in-Aid for Scientific Research on Innovative Areas (26104005).
 
 The electron microscopy was performed at the Division for Medical Research Engineering, Nagoya University Graduate School of Medicine.
 
 A part of this work was conducted at the Nagoya University Nanofabrication Platform, supported by the "Nanotechnology Platform Program" of the Ministry of Education, Culture, Sports, Science and Technology (MEXT), Japan.

 The gelatin used in this paper is offered by Nitta Gelatin Inc.


%

\end{document}